\documentstyle[preprint,aps]{revtex}
\begin{document}
\title{Supersolids in the Bose--Hubbard Hamiltonian}
\author{G.G. Batrouni$^{1}$, R.T. Scalettar$^{2}$,
G.T. Zimanyi$^{2}$, A.P. Kampf$^{\,3}$}
\address{$^{1}$
Thinking Machines Corporation,
245 First Street,
Cambridge, MA 02142}
\address{$^{2}$
Physics Department,
University of California,
Davis, CA 95616}
\address{$^{3}$
Institut f\"ur Theoretische Physik,
Universit\"at zu K\"oln,
50937 K\"oln,
Germany}
\date{\today}
\maketitle
\begin{abstract}
We use a combination of numeric and analytic techniques to determine the ground
state phase diagram of the Bose--Hubbard Hamiltonian with longer range
repulsive
interactions. At half filling one finds superfluidity and an insulating solid
phase. Depending on the relative sizes of near--neighbor and next
near--neighbor interactions, this solid either follows a checkerboard or a
striped pattern. In neither case is there a coexistence with superfluidity.
However upon doping ``supersolid'' phases appear with simultaneous
diagonal and off--diagonal long range order.

\end{abstract}
\newpage
Content-Length: 30634
X-Lines: 647
Status: RO

The interplay between different types of order
in correlated boson and fermion models is a rich area
of current research.
One normally thinks of this interplay as taking the form of a
competition:  In superconducting materials, pairing can compete with
structural phase transitions (A--15's),
charge density wave (CDW) order (Ba$_{1-x}$K$_{x}$BiO$_{3}$),
and antiferromagnetic order (high T$_{c}$ oxides).
This competition is reflected in theoretical models
like the Hubbard Hamiltonian\cite{FHUB} where it is believed that
superconductivity and antiferromagnetism occupy disjoint regions of the phase
diagram, and in electron--phonon models
like the Holstein Hamiltonian\cite{HOLST} where a Peierls--CDW state
usurps pairing at large interaction stengths near half--filling.
Similarly, in bosonic systems with on--site repulsion,
superfluidity competes with an
insulating Mott phase.\cite{FISHER1,BATROUNI}
Nevertheless, it has been suggested that in
$^{4}He$ the solid and superfluid order
might {\it coexist}.\cite{MEISEL,ANDREEV1}
While there are only somewhat indirect experimental
indications for such a phase,\cite{GOODKIND}
``supersolids" have been extensively studied with approximate analytic
approaches.\cite{ANDREEV2,CHESTER,LEGGETT,DZY,FISHERLIU,BRUDER}
There have been to date very few
Monte Carlo simulations.\cite{LOH,OTTER}

In this Letter,
we will explore the question of supersolid order within the context of the
two--dimensional (2D) Bose--Hubbard Hamiltonian\cite{FISHER1}
\begin{equation}
H=-t\sum_{\langle ij \rangle} (a_{i}^{\dagger}a_{j} + a_{j}^{\dagger}a_{i})
  - \mu \sum_{i} n_{i} + V_{0} \sum_{i} n_{i}^{2}
   + V_{1} \sum_{\langle ij \rangle} n_{i}n_{j}
   + V_{2} \sum_{\langle \langle ik \rangle\rangle } n_{i}n_{k}.
\label {eq:eq1}
\end{equation}
Here $a_{i}$ is a boson destruction operator at site $i$, and
$n_{i}=a_{i}^{\dagger}a_{i}$.  The transfer integral $t=1$ sets the scale
of the energy, and $\mu$ is the chemical potential.
$V_{0}, V_{1}$, and $V_{2}$ are on--site,
near--neighbor, and next--near--neighbor
boson--boson repulsions.
In the hard--core limit, the Bose--Hubbard Hamiltonian
maps onto the quantum spin--1/2 Hamiltonian
\begin{equation}
H=-t\sum_{\langle ij \rangle} (S_{i}^{+}S_{j}^{-} + S_{j}^{+}S_{i}^{-})
+V_{1}\sum_{\langle ij \rangle} S_{i}^{z}S_{j}^{z}
+V_{2}\sum_{\langle \langle ik \rangle\rangle } S_{i}^{z}S_{k}^{z}
-H_{z}\sum_{i} S_{i}^{z}.
\label {eq:eq2}
\end{equation}
The field $H_{z}=\mu - 2V_{1}-2V_{2}$.
Ordering of the density corresponds to finite wave vector Ising type order,
whereas superfluidity is described by ferromagnetic ordering in the XY plane.

The mean field phase diagram of Eq.~2
has been worked out by Matsuda and Tsuneto\cite{MATSUDA}.
By increasing the density from half filling the following phases can be
expected:
a Neel state, corresponding to a checkerboard Bose {\it solid} with an ordering
vector $k_{*}=(\pi,\pi)$;
a ferromagnetic phase, with the net moment $M_{xy}\ne 0$ and $M_{z}\ne 0$,
corresponding to a {\it superfluid};
and a fully polarized magnetic phase, where only  $M_{z}\ne 0$,
corresponding to a {\it Mott-insulator}.
As the solid and the superfluid phases possess different broken symmetries,
it is expected that the transition between them is first order.
In an alternative scenario it was proposed that instead there could be two
distinct second order transitions, where the two order parameters
vanish at separate points.\cite{ANDREEV1,CHESTER,LEGGETT}
In the regime between the two transitions {\it both} order parameters
are non-zero, hence it has been termed a {\it supersolid}.
\cite{CHESTER,FISHERLIU,MATSUDA}
The mean field analysis revealed that longer range
forces ($V_{2}$) are needed to stabilize the supersolid,
although recently it was claimed that this conclusion changes in the
soft core case, and
a supersolid phase exists with nearest neighbor interaction alone\cite{STROUD}.
These mean field results need to be reevaluated in the light
of recent studies on the related Heisenberg model with competing
first and second neighbor couplings $J_{1}$ and $J_{2}$
which reveal the possibility of
additional phases:  a collinear phase, with alternating
lines of up and down spins, at large
$J_{2}/J_{1}$\cite{singh,dagotto,larkin}, and a disordered phase
at intermediate values of $J_{2}/J_{1}$\cite{sachdev,kivelson}.

In this paper we address these questions first
by extended mean field and spin wave calculations, then by
Quantum Simulations.

{\bf Mean Field and Spin Wave Analysis.}{\hskip 0.5cm}
We extend the mean field analysis by (i) considering the formation of
a collinear phase, with ordering vector $k_{*}=(\pi, 0)$ or $(0,\pi)$,
and (ii) introducing an approximate soft core representation by allowing
the spin length $S$ to be a variational parameter and adding a term
$H_{constraint}= V_{0} \sum_{i} ({\bf S}_{i}^2 - 1 )^2$ to the Hamiltonian.
We expand the ground state energy around the superfluid phase,
and consider the eigenvalues corresponding to small oscillations of the
density and superfluid order parameter, a procedure equivalent to
generating a Ginzburg-Landau energy. Off half filling
we find three phases: a superfluid, and collinear- and Neel- supersolids.
The phase boundary between the superfluid and the collinear supersolid
is located at $V_{2}= 1+ \delta ^2/[4- \delta ^2 + 24/V_{0}]$,
whereas the phase boundary to the Neel supersolid is at
$V_{2}=V_{1}-1-2\delta ^2/[4- \delta ^2 + 32/V_{0}]$.
Here $\delta=\rho-1/2$, $\rho$ being the density of the boson gas.
The phase diagram for $\rho = 0.53$ and $V_{0}=7$ is shown in Fig.~1.
It displays Neel- and collinear supersolid, and superfluid phases.
At half filling the
supersolid phases vanish, and two insulating solids
are direct neighbors to the superfluid.  This result
is independent of $V_{0}$, {\it ie} it is
true both in the hard and soft core limits.

The analyses of the spin wave fluctuations which exist in the
literature\cite{CHESTER,FISHERLIU,CHENG} are in disagreement.
The spectrum has been found to be either linear\cite{FISHERLIU} or
quadratic\cite{CHESTER,CHENG} {\it at} the density (or field-)
tuned solid-supersolid phase boundary. This dependence is crucial for numerical
studies, as it determines the dynamical critical exponent $z$ and
thereby the appropriate finite size scaling of the lattice.

To settle the issue,
we determine the spin wave spectrum allowing for the formation of a superfluid,
a Neel solid and a Neel supersolid. The calculations involve
a Bogoliubov transformation of coupled density and phase modes.
Details will appear elsewhere.\cite{LONG}
In the Neel solid there are
two excitation branches in a halved Brillouin zone.
Both branches are gapped. In the superfluid there is a Goldstone mode
of linear $k$ dependence at small $k$, and a well developed minimum around
$(\pi,\pi)$.  Taking the continuum limit
identifies this with the roton part of the helium dispersion.
Finally, in the supersolid phase one has a gapless linear mode,
and a gapped one, again in a halved zone.

To clarify the physics of the transitions
we determine the dispersion at the phase boundaries.
At the supersolid-Neel solid transition
the critical mode is at small $k$. For the generic case, $H_{z}\ne 0$,
(no particle-hole symmetry in the boson language), the linear mode
softens into a quadratic one, yielding a dynamical critical exponent $z=2$.
The disappearence of this Goldstone mode signals the destruction of
superfluidity. This value of $z$ agrees with that of
Chester\cite{CHESTER} and Cheng\cite{CHENG}, but differs from that of
Liu and Fisher\cite{FISHERLIU}, who obtain $z=1$.

At the generic superfluid-to-supersolid transition the critical mode is at
$k=k_{*}$: solidification is signalled by the roton mode touching zero.
The rotons also stiffen from a quadratic to a linear minimum, hence $z=1$.
At half filling, i.e. with particle - hole symmetry, both above transitions
have $z=1$.
In a recent Monte Carlo study\cite{OTTER} the same value for $z$
was used in choosing the lattice size to resolve both
the superfluid--supersolid and supersolid--solid transitions, whereas
we find $z=1$ and $z=2$ respectively.
Finally at high fields, at the superfluid-to-Mott
insulator transition the Goldstone mode softens out again, leading to $z=2$,
in agreement with earlier field theoretical predictions\cite{FISHER1}
and numerical simulations\cite{BATROUNI}.


{\bf Quantum Simulations.}{\hskip 0.5cm}
In the remainder of this paper we will discuss the results of Quantum Monte
Carlo (QMC) simulations of the Bose--Hubbard Hamiltonian performed on the
Connection Machine CM5.
We use the world line QMC method in which the 2D quantum partition function
$Z$ of the Bose--Hubbard Hamiltonian is rewritten as a path-integral
over a classical occupation number field $n(\vec j,\tau)$ by
discretizing the inverse temperature $\beta=L_{\tau}\Delta \tau$.
We work in the canonical ensemble, mostly near the special ``half--filled''
point $\rho=N_{b}/N=1/2$.

The presence of solid ordering will be demonstrated by measuring the
equal time density--density correlations,
and their Fourier transform, the structure factor,
\begin{equation}
S(\vec k \,)={1 \over N} \sum_{\vec j \vec l}
e^{i \vec k \, \vec l\,} \langle
n(\vec j,\tau)n(\vec j+\vec l,\tau) \rangle.
\label {eq:eq3}
\end{equation}
Long range solid order in the thermodynamic
limit is signaled by a linear growth of
$S(\vec k_{*})$ with the number of lattice sites, N, at some
ordering vector $\vec k_{*}$.
In Eq.~1, $V_{1}$ drives the formation of a checkerboard
phase with $k_{*}=(\pi,\pi)$, where sites are alternately empty and occupied,
an Ising type Neel antiferromagnet in the spin language.
$V_{2}$ favors a striped phase
where {\it lines} of occupied sites in either the
$x$ or $y$ direction alternate with {\it lines} of empty sites.
In this case the structure factor peaks
at either $\vec k_{*}=(0,\pi)$ or $(\pi,0)$.
We will measure the superfluid density by looking at a topological
property of the boson world lines, the winding number
\cite{BATROUNI,POLLOCK}.

One can determine the ground state phase diagram either by simulating
lattices with large $\beta$ or else by appropriately scaling $L_{\tau} \propto
L^{z}$ with linear spatial lattice size $L$.
The latter technique assumes foreknowledge of $z$ which is
later justified by appropriate scaling behavior,
but has advantages in the precise determination
of phase boundaries.  We shall use it for that purpose when required.
However, as has been done extensively in simulations of both
fermion\cite{WHITE} and boson\cite{BATROUNI} systems, we will primarily
choose $\beta$ large enough so that observables no longer change and we are
assured of measuring ground state properties.

{\bf Results at Half-Filling.}{\hskip 0.5cm}
Our first question is whether the supersolid phase can exist
without a finite $V_{2}$, as suggested by recent work\cite{OTTER,STROUD}.
Fig.~2 shows
the staggered structure factor $S(\pi,\pi)$ and superfluid density
$\rho_{s}$ as a function of near neighbor interaction strength $V_{1}$ at
$V_{0}=7$ and $V_{2}=0$.  We see a sharp transition in both quantities at
$V_{1} \approx 5$.  The raw data already strongly suggest that there is no
supersolid phase intervening between superfluid and solid.  We performed
the appropriate finite size scaling analysis and found that the transition
points differ by at most $0.5\%$, which
we regard as statistically insignificant.
We can now try to drive the supersolid by
turning on the next near neighbor repulsion.  It seems reasonable to
do so near the transition found in Fig.~2.  Therefore, in Fig.~3 we show a
plot of $S(\pi,\pi), S(\pi,0), S(0,\pi)$, and $\rho_{s}$ at $V_{0}=7,
V_{1}=5.5$ ({\it i.e.} just inside the solid phase), sweeping $V_{2}$.  We
see that $V_{2}$ induces a nonzero value of $\rho_{s}$ (only a weak
$V_{2}$ is needed since we started so close to the transition), while
simultaneously destroying the checkerboard order.  For
$V_{2}>V_{1}/2$ we
enter the striped solid phase ($S(\pi,0) \neq 0$),
and again $\rho_{s}$ vanishes.
$S(\pi,\pi)$ also shows a small kink at this superfluid--striped solid
transition.

We have also explored a case of very weak core bosons where $V_{0}=3$.  At
$V_{2}=0$ a transition from SF to checkerboard solid occurs at $V_{1}=6.7$,
with no supersolid in between.  Nor does turning on $V_{2}$ help create one.
Our
conclusion is that, in agreement with our mean field theory, and
contrary to what has recently been found,\cite{OTTER}
no supersolid phase exists at $\rho=1/2$ in the Bose--Hubbard model.

{\bf Supersolids in the Defect Phase.}{\hskip 0.5cm}
We turn next to the doped phase where $\delta=\rho-1/2 \neq 0$.  In Fig.~4
we show a plot of $\rho_s$ and $S(\pi,\pi)$ versus $V_{1}$ at $V_{2}=0$ and
$\delta=0.03$.  We see that a tail of nonzero $\rho_{s}$ persists beyond
the point where the solid has formed.  Fig.~4 contains data for two
lattices sizes at the same doping, demonstrating that the tail is not a
finite size effect.  $\rho_{s}$ drops considerably,
but remains nonzero, as the supersolid is entered.
Our picture is that $N_{b}=N/2$ bosons freeze into a solid, leaving only
the remaining bosons mobile. These then condense into a superfluid.
This is borne out by studying the height of
the tail as a function of $\delta$.  We find that the height is
proportional to $\delta$, indicating that only defect bosons make up the
superfluid condensate within the solid.
We have confirmed that $S(\pi,\pi)$ scales linearly with lattice
size, so that long range crystalline order is indeed present in
the checkerboard supersolid phase.



Fig.~4 exhibits the superfluid in coexistense with the checkerboard solid.
One can also get superfluidity in a striped solid phase.  This is
illustrated in Fig.~5 where we show $S(\pi,0), S(0,\pi)$ and
$\rho_{s}$ at $V_{0}=7, V_{1}=5.5$ and $\delta=0.06$ as a function of
$V_{2}$.  A nonzero value of $\rho_{s}$ is now present in the checkerboard
supersolid at small $V_{2}$.  Increasing $V_{2}$ melts this
supersolid and there is a
large increase in $\rho_{s}$ as all the bosons participate in the
condensate.  At yet larger $V_{2}$ the striped supersolid emerges.
Note that in Fig.~5 we separately plot the
superfluid fraction in the $x$ and $y$ directions.
In the checkerboard supersolid and in the pure superfluid we
find $\rho_{sx}=\rho_{sy}$.  However, in the striped supersolid this
rotational symmetry is broken and $\rho_{sx} \neq \rho_{sy}$.  The symmetry
is broken randomly in the different runs, and there is the expected
correlation between $\rho_{sx}[\rho_{sy}]$, and which of $S(\pi,0)[
S(0,\pi)]$ is large.  {\it The one dimensional superfluid flows only
down the appropriate channels left open by the striped solid phase}.
The boson wavefunction is localized in the orthogonal direction.
In the checkerboard super-solid a similar symmetry breaking occurs,
namely $\rho_{sa} \neq \rho_{sb}$ on the a/b sublattices.\cite{FISHERLIU}.

For our zero
temperature {\it quantum} phase transition, the static periodic potential
of solid bosons does {\it not} confine defects in the Neel
supersolid.  Their wave functions
are still extended Bloch states, from which they can condense into a
superfluid phase.  A
doped boson can move either through an intermediate state
of double occupation of energy cost $\Delta=V_{0}$, or through a move of
two neighboring bosons of energy cost $\Delta=2V_{1}$.  The excess bosons
have extended superfluid wavefunctions with a mass renormalized to
$m_{*}=\sqrt{1 + (\Delta/16t)^{2} } / 2t$ from the bare $m=1/2t$.
That bosons can move in the solid without double occupancy
explains that the supersolid is present in the hard--core limit as well.
Meanwhile, in the striped supersolid, the doped bosons move entirely
freely along the lines of unoccupied sites.
We find $\rho_{s}$ is larger by about a factor of two
in the striped supersolid than in the Neel supersolid at the same doping.

In conclusion, we studied the formation of supersolid phases in interacting
boson systems by mean field, spin-wave
and quantum Monte Carlo techniques.
The results of mean field analysis qualitatively agree with the phase diagram,
obtained by the quantum simulations.
We have shown that a supersolid phase, instead of existing in
some special and narrow window of parameter space, is
a rather generic feature of the
Bose--Hubbard model.\cite{NMU}  Defects or interstitials
introduced into the checkerboard and striped solid phases do not destroy
the diagonal long range order, but rather bose condense into a superfluid.
It is as yet unclear whether this scenario for supersolids
is realized experimentally. There is one positive\cite{GOODKIND} and numerous
negative experiments studying the existence of a supersolid phase in bulk
$^{4}He$ \cite{MEISEL}. Some thin film
studies indicate the existence of superfluidity in incomplete layers
on top of close--packed solid ones.\cite{ANDREEV1} In this situation one
might imagine that {\it different} layers are giving rise to the two types
of order, rather than a single layer being both solid and superfluid.
However, the connection between the theoretical and experimental situations
is not entirely clear, since the precise relation of layer formation and
multiple occupancy in the Bose--Hubbard model is still somewhat
uncertain.\cite{STAIRCASE}

We acknowledge useful discussions with D. Arovas.
This work was supported by National Science Foundation
grant DMR 92-06023 and by Thinking Machines Corporation.
A.P.K. gratefully acknowledges support through a
habilitation scholarship of the Deutsche Forschungsgemeinschaft.

\newpage

{\bf Figure Captions}

Fig.~1:  The mean field phase diagram in the soft-core case.

Fig.~2:  $\rho_{s}$ and $S(\pi,\pi)$ are shown as a function
of $V_{1}$ at $\rho=1/2$, $V_{0}=7$, and $V_{2}=0$.  The lattice size is
8x8x16 and $\tau=1/4$.  There is a single
transition between superfluid and solid.

Fig.~3:  $S(\pi,\pi), S(\pi,0), S(0,\pi)$ and $\rho_{s}$
are shown as a function
of $V_{2}$ at $\rho=1/2$, $V_{0}=7$, and $V_{1}=5.5$.
The lattice is 8x8x16 and $\tau=1/4.$
$V_{2}$ first melts the checkerboard solid into a superfluid,
and then causes a striped solid to form.
$\rho_{s}$ is zero in the solid phases.

Fig.~4:  $\rho_{s}$ and $S(\pi,\pi)$ are shown
as a function of $V_{1}$ at $V_{0}=7.0, V_{2}=0.0,
\tau=1/4$, with the
system doped to $\rho=0.53$.  A superfluid tail persists after the
formation of the checkerboard solid.  Open symbols are
used for $\rho_{s}$ and closed symbols for $S$.  Triangles are
8x8x16 lattices and squares are 10x10x16 lattices.

Fig.~5:  $\rho_{s}$, $S(\pi,0)$ and $S(0,\pi)$ are shown
as a function of $V_{2}$ at $V_{0}=7.0, V_{1}=5.5$,
and the system doped to $\rho_{s}=0.56$.  Here filled and open squares are for
$S(\pi,0)$ and $S(0,\pi)$ respectively.  Crosses and open
triangles show $\rho_{sx}$ and $\rho_{sy}$ respectively.
Both solid phases have nonzero $\rho_{s}$.
In the checkerboard supersolid the
mobile dopants move in 2D (both $\rho_{sx}$ and $\rho_{sy}$ are
nonzero), while here, in the striped phase the superfluid is
confined to the appropriate channels between the lines of
occupied sites.  $S(\pi,\pi)$ is small throughout the parameter range
and is not shown.


\begin{references}
\bibitem{NMU}
Part of the distinction between a supersolid ``window''
and a broader supersolid phase results from the difference between
the nature of the phase diagram in the $N$-$V$ and $\mu$-$V$ planes.
A sweep at constant chemical potential can cut into
a solid phase, revealing a  supersolid window, while
a sweep at fixed density skirts the pure solid
and remains in the supersolid phase.
This occurs because of the existence of a gap in the solid phase,
so that even a slightly doped system may have $\mu$ substantially
shifted from the undoped case.
\end{references}
\end{document}